\documentclass[11pt]{article}
\usepackage{blois}

\usepackage{txfonts}
\usepackage{color}
\usepackage[]{units}

\def\Journal#1#2#3#4{{#1} {\bf #2}, #3 (#4)}

\def\be{\begin{equation}}
\def\ee{\end{equation}}

\newcommand{\hess}{\textsc{H.E.S.S.}}
\newcommand{\fermi}{\textsc{{\it Fermi}-LAT}}

\newcommand{\Photo}{\includegraphics[height=35mm]{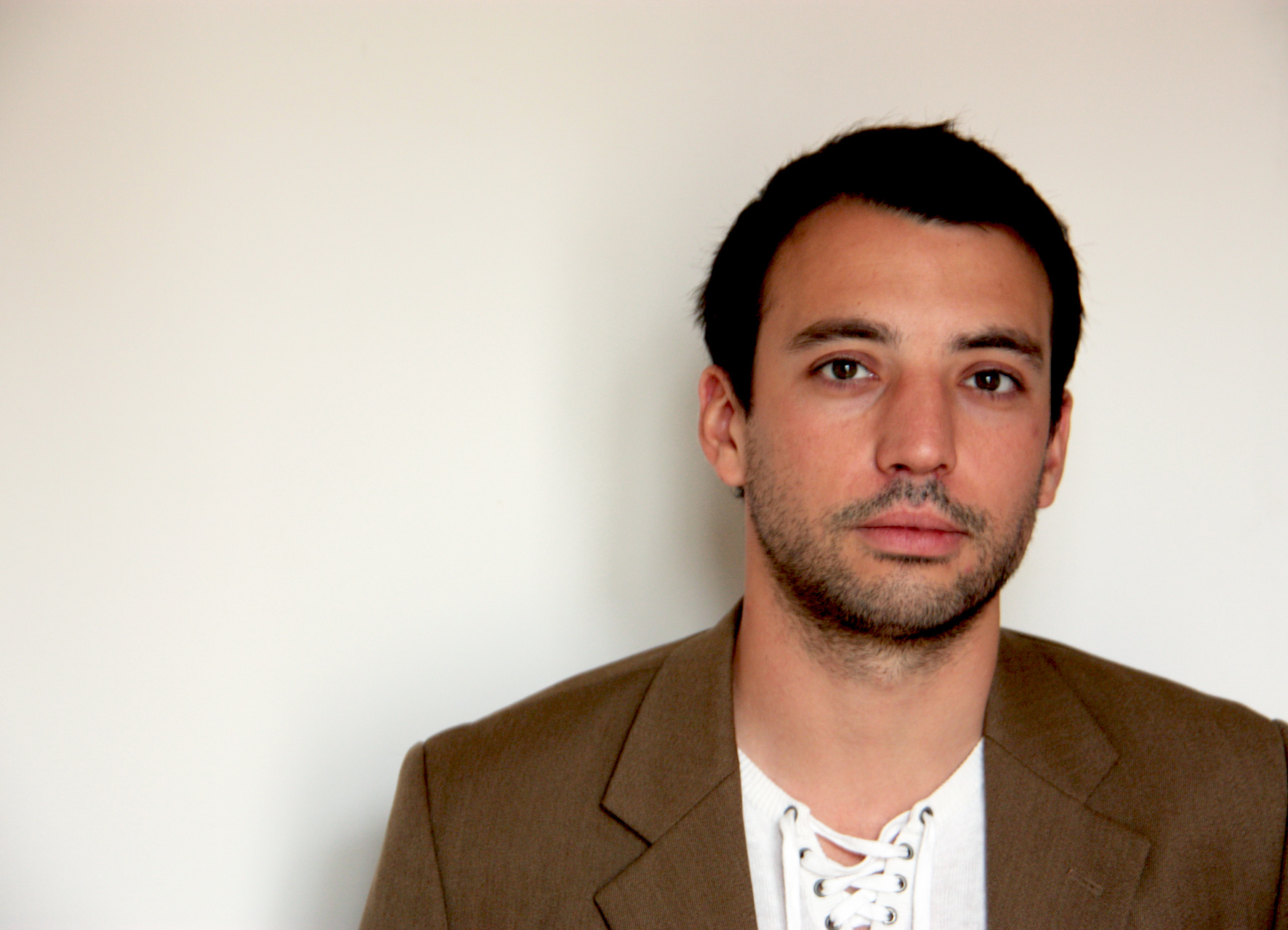}}

\begin{document}
\vspace*{4cm}
\title{Probing the extragalactic background light with H.E.S.S.}

\author{J. BITEAU on behalf of the H.E.S.S. Collaboration}

\address{Laboratoire Leprince-Ringuet, Ecole polytechnique, CNRS/IN2P3, F-91128 Palaiseau, France}

\maketitle\abstracts{The imprint of cosmic backgrounds in the gamma ray spectra of blazars has recently been detected by \hess\ and \fermi, opening the way to studies of gamma-ray propagation on cosmological scales. This proceeding discusses the current constraints on the extragalactic background light (EBL), the effort to increase the collection of blazars detected at TeV energies, and a crucial part of the science case of next-generation instruments: gamma-ray cosmology. }

A new window on the content and history of the universe is being opened by current generation airborne and ground-based gamma-ray instruments. The observational field that is sometimes referred to as {\it gamma-ray cosmology} aims at using the electromagnetic emission of extragalactic sources from tens of GeV up to tens of TeV to constrain diffuse radiations, large scale magnetic fields, dark matter scenarios or fundamental principles such as Lorentz invariance. A first step has been taken with the detection and measurement of the extragalactic background light (EBL) in the near-UV/optical/near-IR bands by gamma-ray instruments such as \hess\ \cite{2013AA...550A...4H} and \fermi\ \cite{2012Sci...338.1190A}.

\section{Gamma rays to probe cosmic backgrounds at micrometric wavelengths}\label{Sec1}

The EBL is the second most intense diffuse radiation in the universe after the Cosmic Microwave Background (CMB). Depending on the wavelength of the EBL photons, one speaks of Cosmic Optical Background (COB), typically from $\unit[0.1]{\mu m}$ to $\unit[10]{\mu m}$, and of Cosmic Infrared Background, from $\unit[10]{\mu m}$ to $\unit[500]{\mu m}$, above which the CMB becomes more intense. The COB includes all the light emitted by stars and galaxies since the end of the cosmic dark ages while the reprocessing by dust of this UV-optical emission in the infrared produces the CIB. 
 
Below hundreds of microns, the EBL is hard to distinguish from our bright local environment, i.e. from emission within the solar system and the Galaxy. The contamination of direct observations by foregrounds results in upper limits on the EBL intensity.  Lower limits are derived with galaxy counts, i.e. cumulating the brightness of galaxies and correcting for the lack of the hardly-detectable faintest ones. Lower bounds on the EBL are matched by models of the local population of galaxies (backward modeling), of large astronomical databases (forward modeling), or semi-analytic models where structure formation is simulated. These various types of modeling predict a COB peak intensity around $\unit[1]{\mu m}$ (where direct measurements exceed galaxy counts by almost an order of magnitude). A comprehensive review of UV-to-IR observations and of the modeling techniques can be found in reference \cite{2001ARAA..39..249H}.

The potential of gamma-ray astronomy for COB measurements was, surprisingly, noticed decades ago. Quoting seminal studies from the late 1960s \cite{1967PhRv..155.1408G}, ``{\it observations of cosmic photons in the region $\it 10^{12}$ to $\it \unit[10^{13}]{\it eV}$ would be of great value, since in this region absorption due to the cosmic optical photons is important.}'' Indeed, for an isotropic target field of photons of energy $E_{\rm EBL}$, the pair-creation cross section peaks when the incoming photon has an energy $E_{\gamma}$ satisfying $E_\gamma E_{\rm EBL}\sim(2 m_e c^2)^2 \sim \unit[1]{MeV^2}$.\cite{1976tper.book.....J} With a peak intensity around $\unit[1]{\mu m}$, corresponding to $E_{\rm EBL}\sim\unit[1]{eV}$, photons from the COB primarily interact with gamma rays of energy $E_\gamma\sim\unit[1]{TeV}$. Still quoting \cite{1967PhRv..155.1408G}, ``{\it this may provide a means of determining the optical photon density and of testing cosmological models}''. Pair creation indeed results in a cosmic opacity to gamma rays that depends both on the cosmological distance of the source and on the EBL density. For a source located at a redshift $z$, the gamma-ray absorption is defined by the optical depth: 

\be
\tau\left(E_{\gamma},z\right) = \int_0^z{\rm d}z' {{{\rm d}l} \over {{\rm d}z}}(z') \int_{0}^{+\infty}{\rm d}E_{\rm EBL}\ n(E_{\rm EBL},z') \int_{-1}^{1}{\rm d}\mu {{1-\mu} \over 2} \sigma_{ee}(E_{\rm EBL},E_{\gamma}\times(1+z'),\mu),
\label{eq:optdepth}
\ee
where the distance element is ${{{\rm d}l} / {{\rm d}z}} = { c / {H_0(1+z)}} { {\sqrt{\Omega_\Lambda + \Omega_m(1+z)^3} } }$ (in a flat universe), $n(E_{\rm EBL},z){\rm d}E_{\rm EBL}$ is the number of EBL photons per unit volume with energies between $E_{\rm EBL}$ and $E_{\rm EBL}+{\rm d}E_{\rm EBL}$, $\mu$ is the cosine of the angle between the momenta of two interacting photons and $\sigma_{ee}$ is the pair creation cross section. Even the observation technique was forecast in the early studies:
``{\it the technique of observing shower Cherenkov radiation would probably be most useful here; however, apparently it can only be used to determine high-energy photon fluxes from discrete sources. Some slight indications that quasars may be such sources has come from observations}''. The reader will note that two decades later the first TeV gamma-ray astrophysical sources were detected and that, almost half a century after the seminal studies, the imprint of the EBL on gamma-ray spectra was finally detected, as discussed in the following.

\section{Detection a a cosmic imprint with TeV gamma rays}\label{Sec2}

The most effective technique to date to detect gamma rays at TeV energies is the imaging of the optical Cherenkov flash emitted in the atmosphere by particles from gamma-ray induced showers. The atmosphere acts as a calorimeter in which the energy of the primary gamma ray can be reconstructed and a stereoscopic view of the shower (using multiple telescopes) helps reconstructing the gamma-ray direction and improves the rejection of background events, such as cosmic proton showers or muons.\cite{Crab}

\begin{figure}
\begin{minipage}{0.49\linewidth}
\centerline{\includegraphics[width=0.9\linewidth]{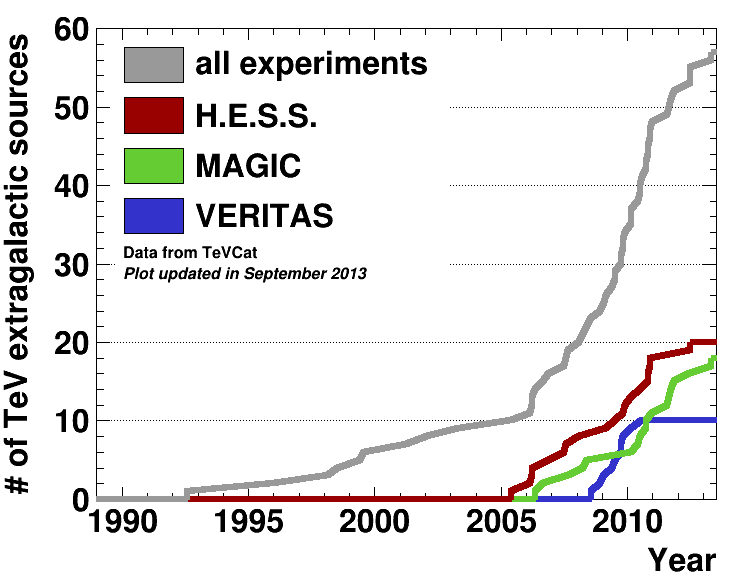}}
\end{minipage}
\hfill
\begin{minipage}{0.49\linewidth}
\centerline{\includegraphics[width=0.9\linewidth]{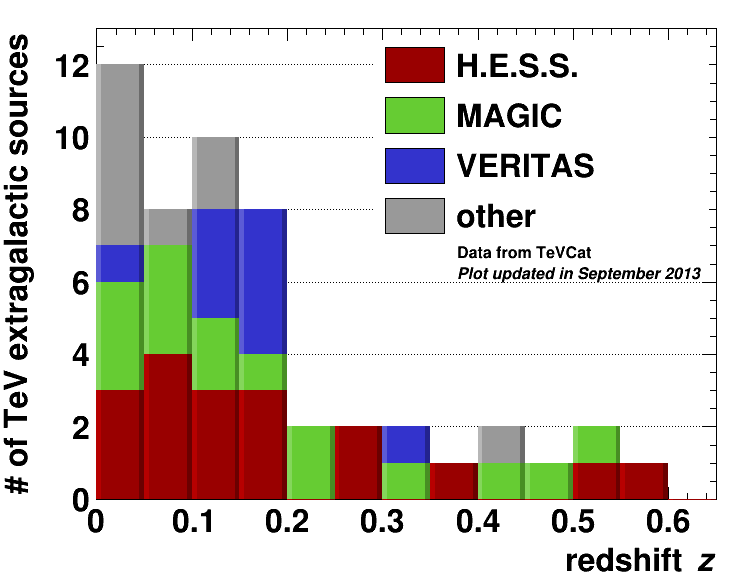}}
\end{minipage}
\caption{Evolution of the number of known extragalactic sources of TeV gamma rays as a function of time (left) and distribution of the redshifts of extragalactic objects with a constrained distance (right). Data from the TeVCat (http://tevcat.in2p3.fr).}
\label{fig:EGAL_status}
\end{figure}

The number of detections of extragalactic sources in TeV gamma-ray has been rapidly growing since the first discovery in 1992.
As shown in Fig.~\ref{fig:EGAL_status} (left), 58 sources of these type have been discovered so far, with a prominent role (48 discoveries) of the current-generation Cherenkov telescopes, ran by the \hess, MAGIC and VERITAS Collaborations. Aside from two nearby starburst galaxies, the extragalactic sky is essentially composed of active galactic nuclei\cite{1993ARAA..31..473A} (AGN, called ``quasars'' in the 70s and 80s), which are heavy power reservoirs fueled by super-massive black holes of billion solar masses. Some of them exhibit jetted structures with opening angles of a few degrees extending from lightyear to million-lightyear scales. The orientation of these two-sided jets in the local universe is random. When a jet is closely aligned with the observer's line of sight, the AGN is called a ``blazar''. With their strongly  beamed emission, blazars account for 90\% of the detected AGN at TeV energies. Most of the detected objects lie at redshifts $z<0.2$ (Fig.~\ref{fig:EGAL_status}, right) because of the quadratic decrease of flux with distance and because of EBL absorption.

Constraining the EBL using gamma rays from blazars was long considered as a typical case of two unknowns (emitted flux and absorption) for a single observable (detected flux).
Stringent limits on the EBL (and on the contribution of primordial stars) have been set fixing the other unknown,\cite{EBLAHA} i.e. assuming a maximum efficiency for the gamma-ray production as characterized by the proportion of gamma rays in the highest-energy band (``hardness'') of the emitted spectrum. The EBL absorption increases as a function of the gamma-ray energy: for a given observed spectrum, the larger the absorption, the harder the emitted spectrum. Assuming a maximum hardness thus results in an upper limit on the EBL density. 

\begin{figure}
\centering
\includegraphics[width=0.62\linewidth]{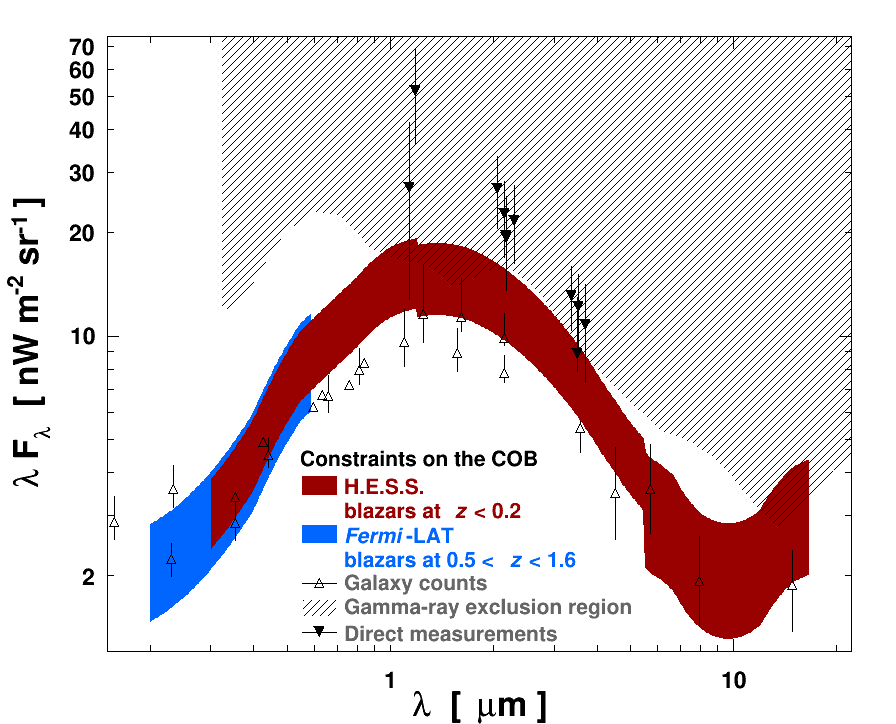}
\caption{The Cosmic Optical Background measured by \hess\ and \fermi\ with blazars at $z<0.2$ and $0.5<z<1.6$, respectively. The $1\sigma$ confidence contours (systematic uncertainties included) correspond to the red and blue filled area.}
\label{fig:EBLstatus}
\end{figure}

The increase of the absorption with energy is actually not steady. Because of the depletion in target photons between $1$ and $\unit[10]{\mu m}$ (decrease in Fig.~\ref{fig:EBLstatus}), the absorption shows an inflection between $1$ and $\unit[10]{TeV}$, which is not expected in the emitted spectrum. This ``cosmic imprint'' acts as the second observable in the problem with two unknowns, enabling a measurement on the EBL absorption. This was studied by the \hess\ Collaboration \cite{2013AA...550A...4H} using the spectra of the seven brightest observed blazars in the southern hemisphere. The emitted fluxes were modeled with smooth functions, i.e. concave or convex functions of energy without inflection point, and a template EBL optical depth \cite{Fran08} scaled by a normalization factor. A combined maximum likelihood technique applied to the datasets from the seven sources shows a normalization factor differing from a null absorption at the $9\sigma$ level, with a best fit value 30\% above the template and statistical and systematic uncertainties of the order of 15\% and 20\%, respectively. The hypotheses of the modeling (EBL template, smooth functions for the emitted flux) prove to be robust and account for less than half of the systematic uncertainties. No outlier is found among the datasets, with a distribution of the individual maximum-likelihood normalizations fully compatible with the statistics. The $1\sigma$ confidence contour derived by the \hess\ Collaboration (red filled area in Fig.~\ref{fig:EBLstatus})  lies in between the lower and upper limits derived from galaxy counts, direct observations and maximum-hardness constraints. The \fermi\ Collaboration similarly detected the EBL absorption at lower energy \cite{2012Sci...338.1190A} using with more distant blazars ($5\sigma$ in the redshift band $0.5<z<1.6$). This is another example of the complementarity of gamma-ray satellites and ground-based telescopes, with \fermi\ probing the near-UV/optical EBL and \hess\ probing the optical and near-infrared bands.

\section{Perspectives of gamma-ray cosmology}\label{Sec3}

The measurements of \hess\ and \fermi\ mark the first firm detections in gamma-ray cosmology. On the theoretical side, an EBL level slightly above the predictions \footnote{Most of the models were initially designed to reproduce lower limits from galaxy counts, making the scaling up expected.} should be reproduced up to about $\unit[5]{\mu m}$. On the experimental/observational side, two important gaps must be filled, the first one being the redshift range $0.2<z<0.5$, where neither \hess\ nor \fermi\ could perform a significant measurement. The second gap lies in between the COB and the CIB, where the emission of polycyclic aromatic hydrocarbon could be dominant. Lowering the energy threshold of ground based instruments, e.g. with \hess~II, will increase the number of detections up to $z=0.5$ and beyond, enabling the study of the evolution of the EBL optical depth.  Above $\unit[5]{\mu m}$, i.e. above $E_\gamma\sim\unit[5]{TeV}$, current measurements remain dominated by statistical uncertainties, pointing out the need for improvements of sensitivity at the highest energies, which will be achieved with the Cherenkov Telescope Array (CTA\cite{CTA}).

Beyond EBL constraints, next steps in gamma-ray cosmology will probably focus on the intergalactic magnetic field (IGMF) filling the voids of the Universe and potentially created in the primordial universe,\cite{IGMF} axion-like particles particles (ALP), a dark matter candidate predicted beyond the Standard Model, and Lorentz Invariance Violation (LIV), studied within Quantum Gravity frameworks. The IGMF could effect pairs produced by EBL absorption of gamma rays and would be tracable either through a broad-band spectral signature or through an apparent morphological extension of the gamma ray source. ALP/gamma-ray oscillations within a magnetized environment could be detected as an erratic spectral modulation in the \hess\ energy range.\cite{ALP} LIV could alter the dispersion relation for gamma rays and impact the energy threshold of pair production, modifying the energy dependence of EBL absorption.\cite{LIV} Studying the propagation of gamma rays on cosmological scales is now feasible and could become one of the major tools shedding light on the components of the Universe and on the laws of modern physics.
  
\section*{Acknowledgments}

Please see standard acknowledgements in \hess\ papers, not reproduced here due to lack of space

\end{document}